\documentclass[prl,twocolumn,superscriptaddress,floatfix,amsmath,amssymb]{revtex4-1}
\usepackage[english]{babel}
\usepackage{graphicx}
\usepackage{dcolumn}
\usepackage{bm}
\usepackage{verbatim}
\usepackage{mathrsfs}
\usepackage{mathtools}
\usepackage{cancel}
\usepackage{epstopdf}
\usepackage{color}
\usepackage[usenames,dvipsnames]{pstricks}
\usepackage{epsfig}
\usepackage[normalem]{ulem}
\usepackage{pst-grad} 
\usepackage{pst-plot} 
\usepackage{hyperref}

\newcommand{\ii}{\mathrm{i}}

\newcommand{\e}{\mathrm{e}}

\newcommand{\nn}{\nonumber}
\newcommand{\be}{\begin{equation}}
\newcommand{\ee}{\end{equation}}
\newcommand{\bea}{\begin{eqnarray}}
\newcommand{\eea}{\end{eqnarray}}
\newcommand{\edu}{\color{magenta}}
\newcommand\numberthis{\addtocounter{equation}{1}\tag{\theequation}}

\def\slashchar#1{\setbox0=\hbox{$#1$} 
\dimen0=\wd0 
\setbox1=\hbox{/} \dimen1=\wd1 
\ifdim\dimen0>\dimen1 
\rlap{\hbox to \dimen0{\hfil/\hfil}} 
#1 
\else 
\rlap{\hbox to \dimen1{\hfil$#1$\hfil}} 
/ 
\fi}

\emergencystretch=\maxdimen
\begin{document}

\title{Quantum Shockwave Communication} 

\author{Aida Ahmadzadegan}
\affiliation{Department of Applied Mathematics, University of Waterloo, Waterloo, Ontario, N2L 3G1, Canada}
\affiliation{Perimeter Institute for Theoretical Physics, Waterloo, Ontario N2L 2Y5, Canada}
\author{Eduardo Martin-Martinez}
\affiliation{Department of Applied Mathematics, University of Waterloo, Waterloo, Ontario, N2L 3G1, Canada}
\affiliation{Perimeter Institute for Theoretical Physics, Waterloo, Ontario N2L 2Y5, Canada}
\affiliation{Institute for Quantum Computing, University of Waterloo, Waterloo, Ontario, N2L 3G1, Canada}
\author{Achim Kempf}
\affiliation{Department of Applied Mathematics, University of Waterloo, Waterloo, Ontario, N2L 3G1, Canada}
\affiliation{Perimeter Institute for Theoretical Physics, Waterloo, Ontario N2L 2Y5, Canada}
\affiliation{Institute for Quantum Computing, University of Waterloo, Waterloo, Ontario, N2L 3G1, Canada}
\affiliation{Department of Physics and Astronomy, University of Waterloo, Waterloo, Ontario N2L 3G1, Canada}

\begin{abstract}

We present a scheme to produce shockwaves in quantum fields by means of pretimed emitters.   
We find that by suitably pre-entangeling the  emitters, the shockwave's energy density can be locally modulated and amplified. When the large amplitudes in such a shockwave are used for communication, the channel capacity depends not only on the signal-to-noise ratio but also on the effect that the entanglement of the emitters has on the correlations in the signal and in the quantum noise at the receiver. As a consequence, by choosing the entanglement of the emitters, the flow of information in the shockwave can be modulated and spatially shaped to some extent independently of the flow of energy. We also 
find that there exists a finite optimal strength of the coupling between the receiver and the quantum field which optimizes the channel capacity by optimizing the tradeoff between sensitivity to the signal and sensitivity to the coupling-induced quantum noise.

\end{abstract}

\maketitle
\setlength{\abovedisplayskip}{4pt}
\setlength{\belowdisplayskip}{4pt}
The phenomenon of shockwaves has been studied in contexts from  the sonic boom to shockwaves in 
superfluids or Bose-Einstein condensates \cite{Damski2004,Dutton663,gurevich1987,QSWfluid}. 
In this Letter, we introduce a type of quantum shockwave that can in principle be created even in empty space. 
To create a shockwave in empty space, we use the fact that, even though no systems can travel faster than the speed of light in vacuum, a suitable line-up of spatially separated quantum emitters, each of which can be at rest, can be pre-timed to successively start emitting into the quantum field in such a way that the sequence of these emitter activations moves superluminally. These emissions can accumulate coherently, creating a shockwave in the quantum field. Such shockwaves are of interest regarding the theoretical limits to the propagation of information through quantum fields and could also possess practical applications, e.g., in quantum communication. 

In particular, we ask if shockwaves produced in this way possess features that have no analog in classical shockwaves. Indeed, we find that by suitably preparing the emitters in an entangled state, it is possible to modulate and thereby to locally enhance the shockwave's energy density and information carrying capacity. Further, we observe that the information flow does not completely track the energy flow. Technically, this is because the energy flow is sensitive only to bi-partite entanglement among the emitters  while the information flow is sensitive to all of their multi-partite entanglement. We also find that, on the receiver's side, the channel capacity is optimized for a specific finite coupling strength between the receiving quantum system and the field. A further increase of the coupling would be detrimental due to a disproportional increase of receiver-induced quantum noise.

\textit{Setup.-} For simplicity, we model the emitters as localized two-level systems, so-called Unruh-DeWitt (UDW) detectors, which can be thought of as simplified descriptions of atoms, molecules or ions \cite{Birrell1984} and we couple these UDW detectors to a scalar quantum field. It is known that this model captures the basic features of the light-matter interaction as long as the exchange of angular momentum does not play a  prominent role, see, e.g., \cite{Martin-Martinez2018,Wavepackets,Pozas2016}. 
Further, we will make use of the idealization that the emitters interact with the field only very briefly. This will allow us to obtain results non-perturbatively. For these methods, see, e.g., \cite{Masahiro2008,Petarnonperturbative}.
Concretely, we consider a free scalar field in $3+1$ dimensional Minkowski spacetime and an array of $n$ generally spacelike separated UDW detectors that will serve as emitters, and one UDW detector that will serve as a receiver. 



We model the interactions of the detectors with the field through the Unruh-DeWitt interaction Hamiltonian \cite{DeWitt,Schlicht2004,Satz2006}. Concretely, 
\bea \label{hamil1}
\!\!\!\!\hat{H}_{\text{I}}^{(i)}(t)&=&\lambda_{(i)} \hat{\mu}_{(i)}(t) \chi_{(i)}(t) \int \mathrm{d}^3 \bm x\, F(\bm x-\bm x_{(i)}) \hat{\phi}(\bm x,t)
\eea
is the interaction Hamiltonian of the $i$-th detector. Its centre of mass is at position $\bm x_{(i)}$ and it interacts with the field for the amount of time determined by the switching function $\chi_{(i)}(t)$. We choose the interaction to be very short by modeling it using a Dirac delta as the switching function, $\chi_{(i)}(t)=\delta(t-t_{(i)})$.
$F(\bm x-\bm x_{(i)})$ is the spatial smearing function of the UDW detector. (Recall that here, in 3+1 dimensions, the interaction of an UDW detector with the field must be smeared in space or time in order to avoid ultraviolet divergences \cite{Schlicht2004,Satz2006}, and that in the light-matter interaction, the smearing functions are given by a product of the atomic wavefunctions of the ground and excited states \cite{Martin-Martinez2018,Pozas2016}.)  
We choose the spatial profile of the detector to be a  sphere of radius $R=1/2$ centred at $\bm{x}_{(i)}=(x_{(i)},y_{(i)},z_{(i)})$:
\bea
\text{sphr}(x\!-\!x_{(i)},y\!-\!y_{(i)},z\!-\!z_{(i)})=
\begin{cases}
1~~~~\text{if}~ |\bm x-\bm x_{(i)}|\leq \frac{1}{2}\\
0 ~~~~\text{otherwise.}
\end{cases}
\eea
Further, $\hat{\mu}_{(i)}(t)$ is the monopole moment operator of the $i$-th detector, $\hat{\mu}_{(i)}(t)=\hat\sigma^{+} e^{\ii \Omega_{(i)} t}+\hat \sigma^{-}e^{-\ii \Omega_{(i)} t}$, where $\Omega_{(i)}$ is the energy gap between the ground state, $\left| g \right\rangle$, and the excited state, $\left|e \right\rangle$, of the detector. Expanding the field in plane-wave modes, \eqref{hamil1} becomes 
\bea
\hat{H}_I^{(i)}(t)&=&\lambda_{(i)} \chi_{(i)}(t)\hat{\mu}_{(i)}(t)\\\nn
&\times&\int \mathrm{d}^3 \bm k \Big(\hat{a}^{\dagger}_{\bm k} \alpha_s(\bm k,\bm x_{(i)},t) +\hat{a}_{\bm k} \alpha_s^*(\bm k,\bm x_{(i)},t) \Big),
\eea
where $\hat a^\dagger_{\bm k}$ and $\hat a_{\bm k}$ are creation and annihilation operators obeying $ [\hat a_{\bm k},\hat a^\dagger_{\bm k'}]=\delta^3(\bm k -\bm k')$ and
\bea \label{alphas}
\alpha_s(\bm k,\bm x_{(i)},t)&=&\int \frac{\mathrm{d}^3 \bm x F(\bm x-\bm x_{(i)}) }{\sqrt{16 \pi^3 \left|\bm k\right|}}\e^{\ii(\left|\bm k\right|t-\bm{k}\cdot\bm{x})}.
\eea

\begin{figure}[htp]
\centering
\vspace{-10pt}
	\includegraphics[width=0.4\textwidth]{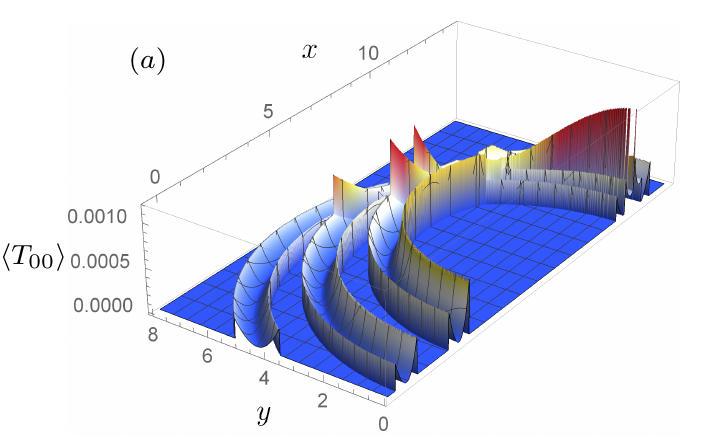}
	\vspace{-5pt}
	\includegraphics[width=0.4\textwidth]{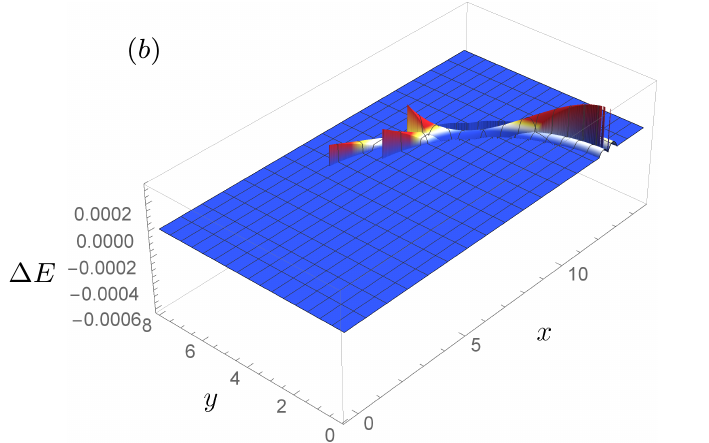}
  \caption{Three entangled emitters are located along the $x$ axis at $x_1=5,~x_2=6.5,~x_3=8$ and couple to the field respectively at $t_1=1,~t_2=2,t_3=3$ with coupling constants $\lambda_1,\lambda_2,\lambda_3=1$ and energy gaps $\Omega_{(i)}=2$. (a) Energy density in the $(x,y)$ plane at time $t=8$ (speed of light $c=1$) when the emitters are initially in the entangled state (\ref{entangled}) with $\theta_1,\theta_2,\theta_3=0$. (b) The difference between the energy density for the entangled initial state shown in (a) and the energy density for the unentangled initial state \eqref{classic}.}\label{stress}
	\end{figure}

\textit{Quantum shockwave}.- Our aim now is to show that, with suitably-timed coupling of the UDW detectors to the field, the field will become excited in the form of a shockwave and that the shockwave is modulated by the choice of the initial entanglement among the emitters. To track the shockwave we first track the energy flow in the field by calculating the expectation value of the energy density $\left\langle \Psi \right|\hat{U}^{\dagger}(t):\hat{T}_{00}:\hat{U}(t)\left|\Psi \right\rangle$, with $\hat{T}_{00}=(\partial_0\hat{\phi})^2+\partial_i \hat{\phi}\partial^i \hat{\phi}$, where for now we assume the total state $\vert \Psi\rangle$ of the emitters and the field to be pure. 
We work in the interaction picture, where the time evolution operator for states is given by 
\be   \label{evop}
\hat{U}(t)=\mathcal{T} \e^{\frac{1}{\ii \hbar}\int_{-\infty}^{\infty}{\hat{H}^{(T)}_I(t') \Theta(t-t')\mathrm{d}t'}},
\ee
where $\mathcal{T}$ represents time ordering and $\hat{H}^{(T)}_I(t')\coloneqq\sum_{i=1}^n\hat{H}_I^{(i)}(t')$. Since the  emitters are chosen to be spacelike separated, their interaction Hamiltonians commute and we can write the time evolution operator as a product of $n$ smeared displacement operators:
\be \label{evolution}
\hat{U}(t)=\prod_{i=1}^n \hat{D}_s^{(i)}(t)=\prod_{i=1}^n \e^{\frac{1}{\ii \hbar}\hat{H}^{(i)}_I(t_{(i)}) \Theta_{(i)}}.
\ee
with $\Theta_{(i)}\coloneqq \Theta(t-t_{(i)})$. To obtain the expectation value of the stress-energy tensor for a general initial state $\left|\Psi \right\rangle$ we use
\vspace{-5pt}
\bea
&\hat{U}^{\dagger}:\hat{T}_{00}:\hat{U}=\hat{D}^{(n)\dagger}_s\dots \hat{D}^{(1)\dagger}_s \hat{T}_{00} \hat{D}^{(1)}_s \dots \hat{D}^{(n)}_s\\\nn
&=\Big(\hat{D}^{(n)\dagger}_s\dots \hat{D}^{(1)\dagger}_s (\partial_0\hat{\phi}(\bm x,t)) \hat{D}^{(1)}_s \dots \hat{D}^{(n)}_s\Big)^2\\\nn
&+\sum_{j=1}^3 \Big(\hat{D}^{(n)\dagger}_s\dots \hat{D}^{(1)\dagger}_s (\partial_j \hat{\phi}(\bm x,t)) \hat{D}^{(1)}_s \dots \hat{D}^{(n)}_s\Big)^2,
\eea
where we used the notation $\hat{U}:=\hat{U}(t)$. We define for $j\in\{0,1,2,3\}$
functions $A(\bm x,t,\bm x_{(i)},t_{(i)})$,
\be
A_{(i),j}(\bm x,t,\bm x_{(i)},t_{(i)})
\coloneqq\int \mathrm{d}^3 \bm k \alpha_{,j}(\bm k, \bm x,t) \alpha_s^*(\bm k,\bm x_{(i)},t_{(i)}).
\ee
so that we then obtain:
\begin{widetext}
\vspace{-8pt}
\bea\label{eqT00}
&\left\langle \Psi\right|\hat{U}^{\dagger}:\hat{T}_{00}:\hat{U}\left|\Psi\right\rangle=\left\langle \Psi\right|\!\Big({\hat{D}^{(n)\dagger}_s}...{\hat{D}^{(1)\dagger}_s} \big(\partial_0 \hat{\phi}(\bm x,t)\big) \hat{D}^{(1)}_s ... \hat{D}^{(n)}_s\Big)^2+\!\sum\limits_{j=1}^3 \Big({\hat{D}^{(n)\dagger}_s}...{\hat{D}^{(1)\dagger}_s} \big(\partial_j \hat{\phi}(\bm x,t)\big) \hat{D}^{(1)}_s ... \hat{D}^{(n)}_s\Big)^2\!\left|\Psi\right\rangle\\\nn
&=\left\langle  \Psi\right|\Big[\big(\partial_0 \hat{\phi}\big)^2+\ii^2 \sum_{i=1}^{n}\lambda^2_{(i)} \hat{\mu}^2_{(i)}\big(t_{(i)}\big)\Theta^2_{(i)}\big(A_{(i)}-A_{(i)}^*\big)^2+2\ii~\partial_0 \hat{\phi} \sum_{i=1}^{n}\lambda_{(i)} \hat{\mu}_{(i)}\big(t_{(i)}\big)\Theta_{(i)}\big(A_{(i)}-A_{(i)}^*\big)\\\nn
&+2\ii^2\sum_{i=1}^{n-1}\sum_{l=i+1}^{n}\lambda_{(i)}\lambda_{(l)} \hat{\mu}_{(i)}\big(t_{(i)}\big)\Theta_{(i)} \hat{\mu}_{(l)}\big(t_{(l)}\big)\Theta_{(l)}\big(A_{(i)}-A_{(i)}^*\big)\big(A_{(l)}-A_{(l)}^*\big)\Big]\left| \Psi\right\rangle\\\nn
&+\left\langle  \Psi\right|\sum\limits_{j=1}^{3}\Big[\big(\partial_j \hat{\phi}\big)^2+\ii^2 \sum\limits_{i=1}^{n}\lambda^2_{(i)} \hat{\mu}^2_{(i)}\big(t_{(i)}\big)\Theta^2_{(i)}\big(A_{(i),j}-A_{(i),j}^*\big)^2+2\ii \partial_j \hat{\phi} \sum\limits_{i=1}^{n}\lambda_{(i)} \hat{\mu}_{(i)}\big(t_{(i)}\big)\Theta_{(i)}\big(A_{(i),j}-A_{(i),j}^*\big)\\\nn
&+2\ii^2\sum\limits_{i=1}^{n-1}\sum\limits_{l=i+1}^{n}\lambda_{(i)}\lambda_{(l)} \hat{\mu}_{(i)}\big(t_{(i)}\big)\Theta_{(i)} \hat{\mu}_{(l)}\big(t_{(l)}\big)\Theta_{(l)}\big(A_{(i),j}-A_{(i),j}^*\big)\big(A_{(l),j}-A_{(l),j}^*\big)\Big]\left| \Psi\right\rangle.
\eea
\vspace{-8pt}
\end{widetext}
For simplicity, for $j\in\{1,2,3\}$, we used the notation $A_{(i),j}\coloneqq A_{(i),j}(\bm x,t,\bm x_{(i)},t_{(i)})$ and $A_{(i)}\coloneqq A_{(i),0}(\bm x,t,\bm x_{(i)},t_{(i)})$. Here, $\alpha_s$ is given in \eqref{alphas} and $\alpha_{,j}(\bm k,\bm x,t)\coloneqq\int \mathrm{d}^3 \bm k \frac{1}{\sqrt{16 \pi^3 \left|\bm k\right|}} \big(\partial_j\e^{\ii(\left|\bm k\right|t-\bm{k}\cdot\bm{x})}\big)$. 
We are now ready to calculate the energy flow of the shockwave and, in particular, to compare the case of unentangled emitters to the case of entangled emitters. 
We consider the example of $n$ emitters that are prepared in a  coherent superposition of states in which one of the detectors is excited, given by the entangled pure state
\be \label{entangled}
\left|W\right\rangle=\frac{1}{\sqrt{n}}\big(\e^{\ii\theta_1}\left|egg...g\right\rangle+\e^{\ii\theta_2}\left|geg...g\right\rangle+...+\e^{\ii\theta_n}\left|g...ge\right\rangle\big).
\ee
The field is in the ground state $\left| 0\right\rangle$ so that $\left| \Psi \right\rangle=\left| W\right\rangle\otimes\left|0\right\rangle$. 
For comparison, we choose the initial state of the emitters to be instead the corresponding incoherent superposition in which one of the detectors is excited, with density matrix: \be \label{classic}
\!\hat \rho=\frac{1}{n}\big(\!\left|egg...\right\rangle\!\left\langle egg...\right|+\left|geg...\right\rangle\!\left\langle geg...\right|+...+\left|...gge\right\rangle\!\left\langle ...gge\right|\!\big)
\ee
The field is again prepared in the ground state so that the total state is now $\hat\rho\otimes\left|0\right\rangle\!\left\langle 0\right|$. In Fig.\ref{stress}(a), we show the start of the buildup of the shockwave for $n=3$ colinear emitters. The spatially extended emitters have spatially extended light cones. Notice that wherever the light cones do not overlap, each light cone displays an energy density distribution which, as expected, is peaked at the leading and trailing edges of the light cone, due to the suddenness of the emitters' coupling to the field.  Fig.\ref{stress}(b) shows a key result, namely that in the region where the emitters'  light cones overlap, i.e., where the shockwave is forming, 
the choice of the emitters' entanglement modulates the energy density. Notice that 
a shockwave also forms for initially unentangled emitters of which only one is excited, as given in \eqref{classic}. The reason is that even a single emitter in its ground state can excite the field, due to the time dependence of its coupling.

\begin{figure}[htp]
 \vspace{-10pt} 
	\includegraphics[width=0.4\textwidth]{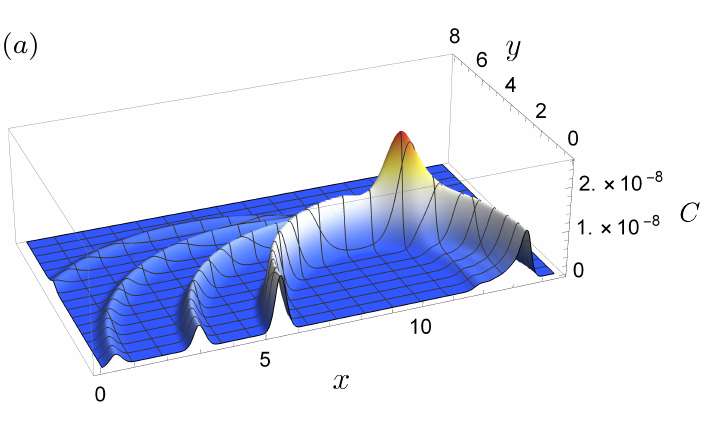}
    \vspace{-10pt}
    \includegraphics[width=0.4\textwidth]{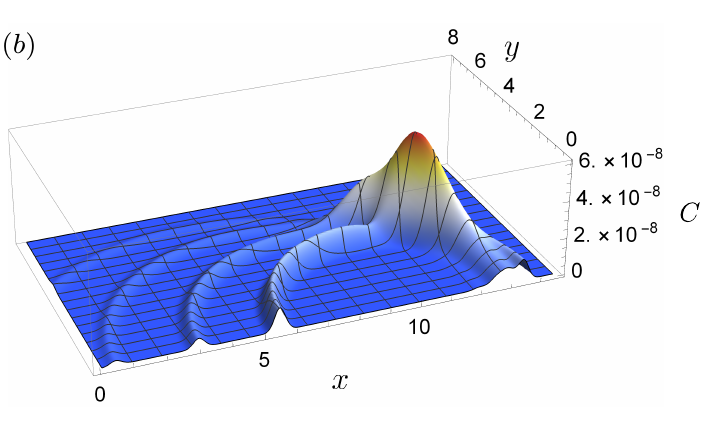}
    	 \vspace{-10pt}
	\includegraphics[width=0.4\textwidth]{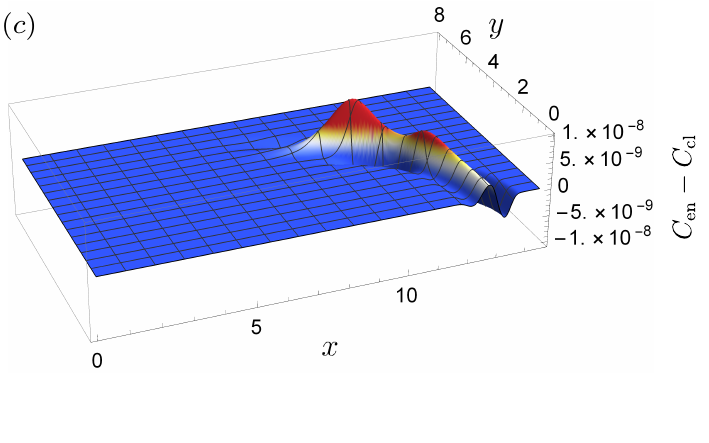}
  \caption{(a) Channel capacity as a function of the location of Bob in the $(x,y)$ plane. Four Alices are prepared in the entangled state 
  \eqref{entangled} with $\theta_1, \theta_2, \theta_3, \theta_4=0$ 
   and located along the $x$ axis at $x_1=5,~x_2=6.5,~x_3=8,~x_4=9.5$. They couple to the field respectively at $t_1=1,~t_2=2,~t_3=3,~t_4=4$ with coupling constants $\lambda_1,\lambda_2,\lambda_3, \lambda_4=0$ or $1$ depending on whether a logical $0$ or $1$ is to be sent. Bob couples to the field at $t_B=8$ with coupling constant $\lambda_B=2$. 
   (b) Same as in (a), except that $\theta_1, \theta_2=0,~\theta_3, \theta_4=\pi$, which is seen to yield a higher peak of the channel capacity. (c) Plot of $\Delta C =C_{en}-C_{cl}$, the difference between the channel capacities when the four Alices start in the state \eqref{entangled} with $\theta_1, \theta_2=0,~\theta_3, \theta_4=\pi$
   versus the corresponding unentangled state \eqref{classic}, showing the quantum enhancement.}\label{entangledwith}
  	\end{figure}

\textit{Channel capacity}.- Let us now calculate the channel capacity for the communication channel between one sender, which consists of a collection of UDW detectors called Alices  (which may or may not be initially entangled) which together emit a quantum shockwave, and one receiver, consisting of an UDW detector called Bob. We choose Bob's detector to have the same spatial profile and energy gap as each of the Alices. We choose the coding scheme such that the Alices collectively send out a logical $``1"$ by means of all Alices coupling to the field so as to create a shockwave, while the Alices encode a logical $``0"$ by not coupling to the field. Bob's initial state 
$\left| \varphi\right\rangle_B$
is the ground state and Bob later registers a logical $``0"$ or $``1"$ by measuring his $\sigma_z$ operator to find it in either the ground or excited state respectively, similar to the scheme used in \cite{Jonsson2015} between one Alice and one Bob.  
To obtain the channel capacity, we first calculate Bob's excitation probability, $p_B(t)$, after the Alices which have been prepared in an entangled pure state, $\left| \varphi\right\rangle_A$, have coupled to the field. To this end, we calculate the expectation value of Bob's observable
$\hat{Q}_B=\left|e\right\rangle\left\langle e\right|$, as a function of space and time, when the total initial state is 
$\left| \Psi\right\rangle=\left| \varphi\right\rangle_A  \left| \varphi\right\rangle_B  \left| 0\right\rangle_f$:
\begin{align*}\label{prob}
&p_B(t)=\left\langle \Psi\right|\hat{U}_A^{\dagger}\hat{U}_B^{\dagger}\hat{Q}_B\hat{U}_A\hat{U}_B\left|\Psi\right\rangle=\left\langle \Psi\right|\frac{\Theta_B}{4}\Big[\hat{B}_0\\
&+\hat{B}\e^{2\ii \lambda_B\hat{\phi}_{s}(\bm k,\bm x_{B},t_{B})}\hat{C}_A+\hat{B}^{\dagger}\e^{-2\ii \lambda_B\hat{\phi}_{s}(\bm k,\bm x_{B},t_{B})}\hat{C}_A^{-1}\Big]\left| \Psi\right\rangle.\numberthis
\end{align*}
Here $\hat{U}_A$ and $\hat{U}_B$ are the contributions to the evolution operator from the Alices and Bob according to \eqref{evolution} and
 
\begin{align}\label{outofline}
&\hat{B}_0\coloneqq \big(\hat{Q}_B+\hat{\mu}_B\hat{Q}_B\hat{\mu}_B\big),\\\nn
&\hat{B}\coloneqq \big[\hat{\mu}_B,\hat{Q}_B\big]+\hat{Q}_B-\hat{\mu}_B\hat{Q}_B\hat{\mu}_B,\\\nn 
&\hat{C}_A\coloneqq \e^{2\lambda_B\!\! \sum\limits_{i=1}^{n} \!\!\lambda_{(i)} \hat{\mu}_{(i)}\Theta_{(i)}\!\int {\!\mathrm{d}^3 \bm k \big( \alpha_s(\bm k,\bm x_{(i)},t_{(i)}) \alpha_s^*(\bm k,\bm x_{B},t_{B})-c.c.\big)}}.
\end{align}
Eq.\eqref{prob} simplifies to $p_B(t)=\left\langle \varphi\right|_A\hat{O}_A\left| \varphi\right\rangle_A$ where

\begin{align*}\label{multipart}
&\hat{O}_A=\frac{\Theta_{B}}{2}\bigg[1-\frac{C_1 }{2}\bigg(\prod_{i=1}^n \Theta_{(i)}\big(\cosh{\gamma_{(i)}}+\hat{\mu}_{(i)}\sinh{\gamma_{(i)}}\big)\\
&+\prod_{i=1}^n \Theta_{(i)}\big(\cosh{\gamma_{(i)}}-\hat{\mu}_{(i)}\sinh{\gamma_{(i)}}\big)\bigg)\bigg].\numberthis
\end{align*}
Here, $\gamma_{(i)}=2\lambda_B\lambda_{(i)}\int \mathrm{d}^3 \bm k \big( \alpha_s(\bm k,\bm x_{(i)},t_{(i)}) \alpha_s^*(\bm k,\bm x_{B},t_{B})-c.c.\big)$. We refer to $p_B(t)$ as $p_e$ if the Alices are initially in an entangled state and then couple to the field (to send a logical 1), and $q_e$ when they do not couple (to send a logical zero). The analogous calculation for the case where the Alices are initially in an unentangled and in this sense classical state yields the probability $p_c$ that Bob gets excited if the Alices couple to the field and $q_c$ when they do not. In both cases, by applying the coding scheme above, we obtain a so-called binary symmetric channel whose channel capacity, $C$, is obtained as \cite{Silvermancapacity}
\bea
C=\frac{-q h(p)+p h(q)}{q-p}+\log_2 \Big[1+2^{\frac{h(p)-h(q)}{q-p}}\Big],
\eea
where $h(x)=-x \log_2[x]-(1-x)\log_2 [1-x]$. In Fig.\ref{entangledwith}(a), we show the example of four Alices prepared in a pure entangled state \eqref{entangled}. The channel capacity between the Alices and Bob is shown as a function of the position of Bob in the $(x,y)$ plane, at a fixed time. Fig.\ref{entangledwith}(b) shows the same as Fig.\ref{entangledwith}(a) but for a different choice of pre-entanglement, demonstrating an enhancement.  
Fig.\ref{entangledwith}(c) shows the difference in the capacity between the case of Fig.\ref{entangledwith}(b) where the Alices are initially entangled  and the case where the Alices start instead in the corresponding incoherent superposition \eqref{classic}. 
The plot demonstrates that entangling the emitters can be used to locally enhance the channel capacity of the shockwave over the case of no pre-entanglement.  
So far, we showed that by entangling the Alices, it is possible to locally modulate both the energy density and the classical channel capacity of the shockwave. 
Further, we observe in equations \eqref{eqT00} and \eqref{multipart}
that the choice of entanglement of the Alices modulates the information flow  differently from the energy flow, i.e., the information flow does not entirely track the energy flow. To see this,  we notice that \eqref{eqT00} contains only products of two monopole operators $\hat\mu_{(i)}$ (ultimately because $\hat{T}_{00}$ is quadratic in $\hat{\phi}$) while \eqref{multipart} contains products of up to all $n$ monopole operators (because the exponentiated Bob-field interaction Hamiltonian from \eqref{evop} contains all powers of the field). This implies that the energy flow is modulated only by bi-partite contributions to the entanglement of the Alices while the information flow is modulated by all multi-partite entanglement among the Alices. We conclude that for senders consisting of multiple pre-entangled emitters such as our Alices, the information flow as measured by the channel capacity possesses a spatial distribution which is richer than what is captured by the ``signal strength" as measured by the energy density. 
 
\begin{figure}[htp]
\centering
	\includegraphics[width=0.35\textwidth]{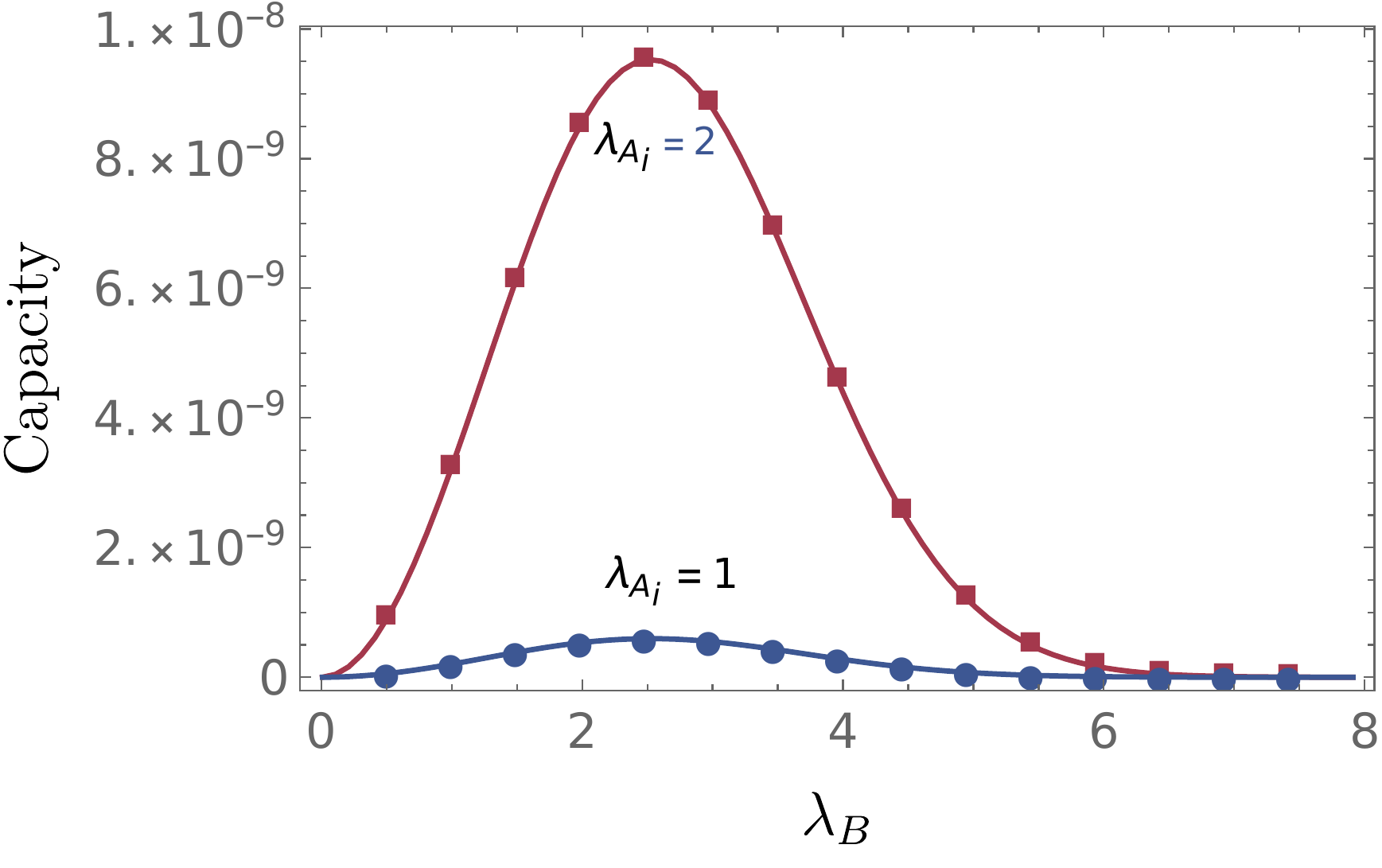}
		\vspace{-5pt}
  \caption{Channel capacity as a function of Bob's coupling strength $\lambda_B$, for two values of coupling strengths of four Alices,  $\lambda_{A_i}=1$ and $\lambda_{A_i}=2$. Bob is located at $x=4.5,y=11$ in the $(x,y)$ plane and the Alices are located along the $x$ axis as in Fig.\ref{entangledwith}. Bob's optimal coupling strength appears to be independent of the Alices' coupling and therefore of the signal strength.}\label{lambdaB}
	\end{figure}

In principle, of course, the sender can always increase both its energy emission and its ability to transmit information not only by recruiting more Alices but also by increasing the effective couplings $\lambda_{(i)}$ of the Alices to the field, as far as an experimental realization allows this. Interestingly, however, when the coupling, $\lambda_{B}$, of the receiver, Bob, to the field is ramped up from zero, i.e., when Bob's receiver is made more and more sensitive to the field and therefore to the Alices' signal then 
the channel capacity increases only initially. Eventually, as $\lambda_{B}$ is increased, Bob's exposure to the noise created by Bob's coupling to the field becomes dominant over Bob's sensitivity to the Alices' signal and the channel capacity starts to drop. This is because, for very large $\lambda_{B}$, Bob's excitation probability after coupling to the field will be close to 0.5 essentially irrespective of the Alices' emissions. 
Indeed, as shown in Fig.\ref{lambdaB}, our nonperturbative calculation demonstrates that there exists a finite optimal coupling strength for Bob that maximizes the channel capacity. 
	
\textit{Outlook.-}
The present setup generalizes to the creation of shockwaves in any quantum field in any medium and, in principle, even to the generating of gravitational shockwaves. While it is not possible to shield gravity, individual masses (that may be pre-entangled) could be moved with suitably prearranged timing. A very interesting challenge will also be the study of quantum channel capacities for quantum shockwaves, or their capacity to transmit pre-existing entanglement with ancillas.  
It will also be interesting to determine how the coupling strength $\lambda^{(opt)}_B$ that optimizes the channel capacity depends on Bob's switching and smearing functions as well as on Bob's energy gap. Also, it remains to determine the optimal choice of the Alices' entanglement for maximizing the energy density or the channel capacity, respectively. Notice that the quantum shockwaves here also admit entirely nonclassical coding schemes that encode logical bits not as usual, e.g., by amplitude of frequency modulation but instead by entanglement modulation, i.e., by varying exclusively the pre-entanglement of the Alices. Further, when operated in time reversal, the quantum shockwave scheme may be useful for quantum measurements such as superresolution, e.g., for telescopes or microscopes, essentially by using pre-entangled systems that play the role of camera pixels whose sensitivity window is suitably timed. In fact, our setup here naturally generalizes to schemes in which the sender and the receiver recruit not only multiple Alices but also multiple Bobs, in analogy to classical MIMO wireless systems which are today in ubiquitous use \cite{mimo1970,mimo1974,mimoreview2002,mimoreview2004,mimoreview2011}. While classical MIMO allows one to increase the capacity by using phases, the quantum setup will allow further modulation and enhancement through entanglement, which may also be useful for quantum cryptographic purposes. Experiments on the new type of quantum shockwave may already be possible using superconducting qubits coupled to microwave transmission lines. There, a potentially sufficiently high degree of control has been achieved in the ability to couple quantum emitters non-locally \cite{Goran}, as well as very fast and ultra-strongly \cite{Peropadre,FornNature}. In particular, we here used Dirac delta switching functions which enabled us to obtain exact results, non-perturbatively. Indeed, for superconducting qubits, the regime of delta switching functions can be approached since the interactions can be varied at scales that are much faster than the inverse of the qubit frequency~\cite{Deng_2015_Floquet}.

\acknowledgements

\textit{Acknowledgments.-}
AK and EM-M acknowledge support through the Discovery Grant Program of the Canadian National Science and Engineering Research Council (NSERC). EM-M is also partially supported by an Ontario Early Researcher Award.

\bibliography{cavity_refs}

\begin{thebibliography}{24}%
\makeatletter
\providecommand \@ifxundefined [1]{%
 \@ifx{#1\undefined}
}%
\providecommand \@ifnum [1]{%
 \ifnum #1\expandafter \@firstoftwo
 \else \expandafter \@secondoftwo
 \fi
}%
\providecommand \@ifx [1]{%
 \ifx #1\expandafter \@firstoftwo
 \else \expandafter \@secondoftwo
 \fi
}%
\providecommand \natexlab [1]{#1}%
\providecommand \enquote  [1]{``#1''}%
\providecommand \bibnamefont  [1]{#1}%
\providecommand \bibfnamefont [1]{#1}%
\providecommand \citenamefont [1]{#1}%
\providecommand \href@noop [0]{\@secondoftwo}%
\providecommand \href [0]{\begingroup \@sanitize@url \@href}%
\providecommand \@href[1]{\@@startlink{#1}\@@href}%
\providecommand \@@href[1]{\endgroup#1\@@endlink}%
\providecommand \@sanitize@url [0]{\catcode `\\12\catcode `\$12\catcode
  `\&12\catcode `\#12\catcode `\^12\catcode `\_12\catcode `\%12\relax}%
\providecommand \@@startlink[1]{}%
\providecommand \@@endlink[0]{}%
\providecommand \url  [0]{\begingroup\@sanitize@url \@url }%
\providecommand \@url [1]{\endgroup\@href {#1}{\urlprefix }}%
\providecommand \urlprefix  [0]{URL }%
\providecommand \Eprint [0]{\href }%
\providecommand \doibase [0]{http://dx.doi.org/}%
\providecommand \selectlanguage [0]{\@gobble}%
\providecommand \bibinfo  [0]{\@secondoftwo}%
\providecommand \bibfield  [0]{\@secondoftwo}%
\providecommand \translation [1]{[#1]}%
\providecommand \BibitemOpen [0]{}%
\providecommand \bibitemStop [0]{}%
\providecommand \bibitemNoStop [0]{.\EOS\space}%
\providecommand \EOS [0]{\spacefactor3000\relax}%
\providecommand \BibitemShut  [1]{\csname bibitem#1\endcsname}%
\let\auto@bib@innerbib\@empty
\bibitem [{\citenamefont {Damski}(2004)}]{Damski2004}%
  \BibitemOpen
  \bibfield  {author} {\bibinfo {author} {\bibfnamefont {B.}~\bibnamefont
  {Damski}},\ }\href {\doibase 10.1103/PhysRevA.69.043610} {\bibfield
  {journal} {\bibinfo  {journal} {Phys. Rev. A}\ }\textbf {\bibinfo {volume}
  {69}},\ \bibinfo {pages} {043610} (\bibinfo {year} {2004})}\BibitemShut
  {NoStop}%
\bibitem [{\citenamefont {Dutton}\ \emph {et~al.}(2001)\citenamefont {Dutton},
  \citenamefont {Budde}, \citenamefont {Slowe},\ and\ \citenamefont
  {Hau}}]{Dutton663}%
  \BibitemOpen
  \bibfield  {author} {\bibinfo {author} {\bibfnamefont {Z.}~\bibnamefont
  {Dutton}}, \bibinfo {author} {\bibfnamefont {M.}~\bibnamefont {Budde}},
  \bibinfo {author} {\bibfnamefont {C.}~\bibnamefont {Slowe}}, \ and\ \bibinfo
  {author} {\bibfnamefont {L.~V.}\ \bibnamefont {Hau}},\ }\href {\doibase
  10.1126/science.1062527} {\bibfield  {journal} {\bibinfo  {journal}
  {Science}\ }\textbf {\bibinfo {volume} {293}},\ \bibinfo {pages} {663}
  (\bibinfo {year} {2001})}\BibitemShut {NoStop}%
\bibitem [{\citenamefont {Gurevich}\ and\ \citenamefont
  {Krylov}(1987)}]{gurevich1987}%
  \BibitemOpen
  \bibfield  {author} {\bibinfo {author} {\bibfnamefont {A.}~\bibnamefont
  {Gurevich}}\ and\ \bibinfo {author} {\bibfnamefont {A.}~\bibnamefont
  {Krylov}},\ }\href@noop {} {\bibfield  {journal} {\bibinfo  {journal} {Zh.
  Eksp. Teor. Fiz}\ }\textbf {\bibinfo {volume} {92}},\ \bibinfo {pages} {1684}
  (\bibinfo {year} {1987})}\BibitemShut {NoStop}%
\bibitem [{\citenamefont {Bulgac}\ \emph {et~al.}(2012)\citenamefont {Bulgac},
  \citenamefont {Luo},\ and\ \citenamefont {Roche}}]{QSWfluid}%
  \BibitemOpen
  \bibfield  {author} {\bibinfo {author} {\bibfnamefont {A.}~\bibnamefont
  {Bulgac}}, \bibinfo {author} {\bibfnamefont {Y.-L.}\ \bibnamefont {Luo}}, \
  and\ \bibinfo {author} {\bibfnamefont {K.~J.}\ \bibnamefont {Roche}},\ }\href
  {\doibase 10.1103/PhysRevLett.108.150401} {\bibfield  {journal} {\bibinfo
  {journal} {Phys. Rev. Lett.}\ }\textbf {\bibinfo {volume} {108}},\ \bibinfo
  {pages} {150401} (\bibinfo {year} {2012})}\BibitemShut {NoStop}%
\bibitem [{\citenamefont {Birrell}\ and\ \citenamefont
  {Davies}(1984)}]{Birrell1984}%
  \BibitemOpen
  \bibfield  {author} {\bibinfo {author} {\bibfnamefont {N.}~\bibnamefont
  {Birrell}}\ and\ \bibinfo {author} {\bibfnamefont {P.}~\bibnamefont
  {Davies}},\ }\href@noop {} {\emph {\bibinfo {title} {Quantum fields in curved
  space}}}\ (\bibinfo  {publisher} {Cambridge university press. England.},\
  \bibinfo {year} {1984})\BibitemShut {NoStop}%
\bibitem [{\citenamefont {Mart\'{i}n-Mart\'{i}nez}\ and\ \citenamefont
  {Rodriguez-Lopez}(2018)}]{Martin-Martinez2018}%
  \BibitemOpen
  \bibfield  {author} {\bibinfo {author} {\bibfnamefont {E.}~\bibnamefont
  {Mart\'{i}n-Mart\'{i}nez}}\ and\ \bibinfo {author} {\bibfnamefont
  {P.}~\bibnamefont {Rodriguez-Lopez}},\ }\href {\doibase
  10.1103/PhysRevD.97.105026} {\bibfield  {journal} {\bibinfo  {journal} {Phys.
  Rev. D}\ }\textbf {\bibinfo {volume} {97}},\ \bibinfo {pages} {105026}
  (\bibinfo {year} {2018})}\BibitemShut {NoStop}%
\bibitem [{\citenamefont {Mart\'{i}n-Mart\'{i}nez}\ \emph
  {et~al.}(2013)\citenamefont {Mart\'{i}n-Mart\'{i}nez}, \citenamefont
  {Montero},\ and\ \citenamefont {del Rey}}]{Wavepackets}%
  \BibitemOpen
  \bibfield  {author} {\bibinfo {author} {\bibfnamefont {E.}~\bibnamefont
  {Mart\'{i}n-Mart\'{i}nez}}, \bibinfo {author} {\bibfnamefont
  {M.}~\bibnamefont {Montero}}, \ and\ \bibinfo {author} {\bibfnamefont
  {M.}~\bibnamefont {del Rey}},\ }\href {\doibase 10.1103/PhysRevD.87.064038}
  {\bibfield  {journal} {\bibinfo  {journal} {Phys. Rev. D}\ }\textbf {\bibinfo
  {volume} {87}},\ \bibinfo {pages} {064038} (\bibinfo {year}
  {2013})}\BibitemShut {NoStop}%
\bibitem [{\citenamefont {Pozas-Kerstjens}\ and\ \citenamefont
  {Mart\'{\i}n-Mart\'{\i}nez}(2016)}]{Pozas2016}%
  \BibitemOpen
  \bibfield  {author} {\bibinfo {author} {\bibfnamefont {A.}~\bibnamefont
  {Pozas-Kerstjens}}\ and\ \bibinfo {author} {\bibfnamefont {E.}~\bibnamefont
  {Mart\'{\i}n-Mart\'{\i}nez}},\ }\href {\doibase 10.1103/PhysRevD.94.064074}
  {\bibfield  {journal} {\bibinfo  {journal} {Phys. Rev. D}\ }\textbf {\bibinfo
  {volume} {94}},\ \bibinfo {pages} {064074} (\bibinfo {year}
  {2016})}\BibitemShut {NoStop}%
\bibitem [{\citenamefont {Hotta}(2008)}]{Masahiro2008}%
  \BibitemOpen
  \bibfield  {author} {\bibinfo {author} {\bibfnamefont {M.}~\bibnamefont
  {Hotta}},\ }\href {\doibase 10.1103/PhysRevD.78.045006} {\bibfield  {journal}
  {\bibinfo  {journal} {Phys. Rev. D}\ }\textbf {\bibinfo {volume} {78}},\
  \bibinfo {pages} {045006} (\bibinfo {year} {2008})}\BibitemShut {NoStop}%
\bibitem [{\citenamefont {Simidzija}\ and\ \citenamefont
  {Mart\'{\i}n-Mart\'{\i}nez}(2017)}]{Petarnonperturbative}%
  \BibitemOpen
  \bibfield  {author} {\bibinfo {author} {\bibfnamefont {P.}~\bibnamefont
  {Simidzija}}\ and\ \bibinfo {author} {\bibfnamefont {E.}~\bibnamefont
  {Mart\'{\i}n-Mart\'{\i}nez}},\ }\href {\doibase 10.1103/PhysRevD.96.065008}
  {\bibfield  {journal} {\bibinfo  {journal} {Phys. Rev. D}\ }\textbf {\bibinfo
  {volume} {96}},\ \bibinfo {pages} {065008} (\bibinfo {year}
  {2017})}\BibitemShut {NoStop}%
\bibitem [{\citenamefont {DeWitt}(1980)}]{DeWitt}%
  \BibitemOpen
  \bibfield  {author} {\bibinfo {author} {\bibfnamefont {B.}~\bibnamefont
  {DeWitt}},\ }\href@noop {} {\emph {\bibinfo {title} {General Relativity; an
  Einstein Centenary Survey}}}\ (\bibinfo  {publisher} {Cambridge University
  Press},\ \bibinfo {address} {Cambridge, UK},\ \bibinfo {year}
  {1980})\BibitemShut {NoStop}%
\bibitem [{\citenamefont {Schlicht}(2004)}]{Schlicht2004}%
  \BibitemOpen
  \bibfield  {author} {\bibinfo {author} {\bibfnamefont {S.}~\bibnamefont
  {Schlicht}},\ }\href@noop {} {\bibfield  {journal} {\bibinfo  {journal}
  {Class. Quant. Grav.}\ }\textbf {\bibinfo {volume} {21}},\ \bibinfo {pages}
  {4647} (\bibinfo {year} {2004})}\BibitemShut {NoStop}%
\bibitem [{\citenamefont {Louko}\ and\ \citenamefont {Satz}(2006)}]{Satz2006}%
  \BibitemOpen
  \bibfield  {author} {\bibinfo {author} {\bibfnamefont {J.}~\bibnamefont
  {Louko}}\ and\ \bibinfo {author} {\bibfnamefont {A.}~\bibnamefont {Satz}},\
  }\href@noop {} {\bibfield  {journal} {\bibinfo  {journal} {Class. Quant.
  Grav.}\ }\textbf {\bibinfo {volume} {23}},\ \bibinfo {pages} {6321} (\bibinfo
  {year} {2006})}\BibitemShut {NoStop}%
\bibitem [{\citenamefont {Jonsson}\ \emph {et~al.}(2015)\citenamefont
  {Jonsson}, \citenamefont {Mart\'{\i}n-Mart\'{\i}nez},\ and\ \citenamefont
  {Kempf}}]{Jonsson2015}%
  \BibitemOpen
  \bibfield  {author} {\bibinfo {author} {\bibfnamefont {R.~H.}\ \bibnamefont
  {Jonsson}}, \bibinfo {author} {\bibfnamefont {E.}~\bibnamefont
  {Mart\'{\i}n-Mart\'{\i}nez}}, \ and\ \bibinfo {author} {\bibfnamefont
  {A.}~\bibnamefont {Kempf}},\ }\href {\doibase 10.1103/PhysRevLett.114.110505}
  {\bibfield  {journal} {\bibinfo  {journal} {Phys. Rev. Lett.}\ }\textbf
  {\bibinfo {volume} {114}},\ \bibinfo {pages} {110505} (\bibinfo {year}
  {2015})}\BibitemShut {NoStop}%
\bibitem [{\citenamefont {Silverman}(1955)}]{Silvermancapacity}%
  \BibitemOpen
  \bibfield  {author} {\bibinfo {author} {\bibfnamefont {R.}~\bibnamefont
  {Silverman}},\ }\href {\doibase 10.1109/TIT.1955.1055138} {\bibfield
  {journal} {\bibinfo  {journal} {IRE Transactions on Information Theory}\
  }\textbf {\bibinfo {volume} {1}},\ \bibinfo {pages} {19} (\bibinfo {year}
  {1955})}\BibitemShut {NoStop}%
\bibitem [{\citenamefont {Kaye}\ and\ \citenamefont {George}(1970)}]{mimo1970}%
  \BibitemOpen
  \bibfield  {author} {\bibinfo {author} {\bibfnamefont {A.}~\bibnamefont
  {Kaye}}\ and\ \bibinfo {author} {\bibfnamefont {D.}~\bibnamefont {George}},\
  }\href {\doibase 10.1109/TCOM.1970.1090417} {\bibfield  {journal} {\bibinfo
  {journal} {IEEE Trans. Commun. Technol.}\ }\textbf {\bibinfo {volume} {18}},\
  \bibinfo {pages} {520} (\bibinfo {year} {1970})}\BibitemShut {NoStop}%
\bibitem [{\citenamefont {Brandenburg}\ and\ \citenamefont
  {Wyner}(1974)}]{mimo1974}%
  \BibitemOpen
  \bibfield  {author} {\bibinfo {author} {\bibfnamefont {L.~H.}\ \bibnamefont
  {Brandenburg}}\ and\ \bibinfo {author} {\bibfnamefont {A.~D.}\ \bibnamefont
  {Wyner}},\ }\href {\doibase 10.1002/j.1538-7305.1974.tb02768.x} {\bibfield
  {journal} {\bibinfo  {journal} {The Bell System Technical Journal}\ }\textbf
  {\bibinfo {volume} {53}},\ \bibinfo {pages} {745} (\bibinfo {year}
  {1974})}\BibitemShut {NoStop}%
\bibitem [{\citenamefont {Yu}\ and\ \citenamefont
  {Ottersten}()}]{mimoreview2002}%
  \BibitemOpen
  \bibfield  {author} {\bibinfo {author} {\bibfnamefont {K.}~\bibnamefont
  {Yu}}\ and\ \bibinfo {author} {\bibfnamefont {B.}~\bibnamefont {Ottersten}},\
  }\href {\doibase 10.1002/wcm.78} {\bibfield  {journal} {\bibinfo  {journal}
  {Wireless Communications and Mobile Computing}\ }\textbf {\bibinfo {volume}
  {2}},\ \bibinfo {pages} {653}}\BibitemShut {NoStop}%
\bibitem [{\citenamefont {Jensen}\ and\ \citenamefont
  {Wallace}(2004)}]{mimoreview2004}%
  \BibitemOpen
  \bibfield  {author} {\bibinfo {author} {\bibfnamefont {M.~A.}\ \bibnamefont
  {Jensen}}\ and\ \bibinfo {author} {\bibfnamefont {J.~W.}\ \bibnamefont
  {Wallace}},\ }\href {\doibase 10.1109/TAP.2004.835272} {\bibfield  {journal}
  {\bibinfo  {journal} {IEEE Trans. Antennas Propag.}\ }\textbf {\bibinfo
  {volume} {52}},\ \bibinfo {pages} {2810} (\bibinfo {year}
  {2004})}\BibitemShut {NoStop}%
\bibitem [{\citenamefont {Arapoglou}\ \emph {et~al.}(2011)\citenamefont
  {Arapoglou}, \citenamefont {Liolis}, \citenamefont {Bertinelli},
  \citenamefont {Panagopoulos}, \citenamefont {Cottis},\ and\ \citenamefont
  {Gaudenzi}}]{mimoreview2011}%
  \BibitemOpen
  \bibfield  {author} {\bibinfo {author} {\bibfnamefont {P.}~\bibnamefont
  {Arapoglou}}, \bibinfo {author} {\bibfnamefont {K.}~\bibnamefont {Liolis}},
  \bibinfo {author} {\bibfnamefont {M.}~\bibnamefont {Bertinelli}}, \bibinfo
  {author} {\bibfnamefont {A.}~\bibnamefont {Panagopoulos}}, \bibinfo {author}
  {\bibfnamefont {P.}~\bibnamefont {Cottis}}, \ and\ \bibinfo {author}
  {\bibfnamefont {R.~D.}\ \bibnamefont {Gaudenzi}},\ }\href {\doibase
  10.1109/SURV.2011.033110.00072} {\bibfield  {journal} {\bibinfo  {journal}
  {IEEE Commun. Surveys Tuts.}\ }\textbf {\bibinfo {volume} {13}},\ \bibinfo
  {pages} {27} (\bibinfo {year} {2011})}\BibitemShut {NoStop}%
\bibitem [{\citenamefont {Guo}\ \emph {et~al.}(2017)\citenamefont {Guo},
  \citenamefont {Grimsmo}, \citenamefont {Kockum}, \citenamefont {Pletyukhov},\
  and\ \citenamefont {Johansson}}]{Goran}%
  \BibitemOpen
  \bibfield  {author} {\bibinfo {author} {\bibfnamefont {L.}~\bibnamefont
  {Guo}}, \bibinfo {author} {\bibfnamefont {A.}~\bibnamefont {Grimsmo}},
  \bibinfo {author} {\bibfnamefont {A.~F.}\ \bibnamefont {Kockum}}, \bibinfo
  {author} {\bibfnamefont {M.}~\bibnamefont {Pletyukhov}}, \ and\ \bibinfo
  {author} {\bibfnamefont {G.}~\bibnamefont {Johansson}},\ }\href {\doibase
  10.1103/PhysRevA.95.053821} {\bibfield  {journal} {\bibinfo  {journal} {Phys.
  Rev. A}\ }\textbf {\bibinfo {volume} {95}},\ \bibinfo {pages} {053821}
  (\bibinfo {year} {2017})}\BibitemShut {NoStop}%
\bibitem [{\citenamefont {Peropadre}\ \emph {et~al.}(2010)\citenamefont
  {Peropadre}, \citenamefont {Forn-D\'{\i}az}, \citenamefont {Solano},\ and\
  \citenamefont {Garc\'{\i}a-Ripoll}}]{Peropadre}%
  \BibitemOpen
  \bibfield  {author} {\bibinfo {author} {\bibfnamefont {B.}~\bibnamefont
  {Peropadre}}, \bibinfo {author} {\bibfnamefont {P.}~\bibnamefont
  {Forn-D\'{\i}az}}, \bibinfo {author} {\bibfnamefont {E.}~\bibnamefont
  {Solano}}, \ and\ \bibinfo {author} {\bibfnamefont {J.~J.}\ \bibnamefont
  {Garc\'{\i}a-Ripoll}},\ }\href {\doibase 10.1103/PhysRevLett.105.023601}
  {\bibfield  {journal} {\bibinfo  {journal} {Phys. Rev. Lett.}\ }\textbf
  {\bibinfo {volume} {105}},\ \bibinfo {pages} {023601} (\bibinfo {year}
  {2010})}\BibitemShut {NoStop}%
\bibitem [{\citenamefont {Forn-D{\'\i}az}\ \emph {et~al.}(2017)\citenamefont
  {Forn-D{\'\i}az}, \citenamefont {Garc{\'\i}a-Ripoll}, \citenamefont
  {Peropadre}, \citenamefont {Orgiazzi}, \citenamefont {Yurtalan},
  \citenamefont {Belyansky}, \citenamefont {Wilson},\ and\ \citenamefont
  {Lupascu}}]{FornNature}%
  \BibitemOpen
  \bibfield  {author} {\bibinfo {author} {\bibfnamefont {P.}~\bibnamefont
  {Forn-D{\'\i}az}}, \bibinfo {author} {\bibfnamefont {J.~J.}\ \bibnamefont
  {Garc{\'\i}a-Ripoll}}, \bibinfo {author} {\bibfnamefont {B.}~\bibnamefont
  {Peropadre}}, \bibinfo {author} {\bibfnamefont {J.-L.}\ \bibnamefont
  {Orgiazzi}}, \bibinfo {author} {\bibfnamefont {M.~A.}\ \bibnamefont
  {Yurtalan}}, \bibinfo {author} {\bibfnamefont {R.}~\bibnamefont {Belyansky}},
  \bibinfo {author} {\bibfnamefont {C.~M.}\ \bibnamefont {Wilson}}, \ and\
  \bibinfo {author} {\bibfnamefont {A.}~\bibnamefont {Lupascu}},\ }\href
  {\doibase 10.1038/nphys3905} {\bibfield  {journal} {\bibinfo  {journal}
  {Nature Physics}\ }\textbf {\bibinfo {volume} {13}},\ \bibinfo {pages} {39}
  (\bibinfo {year} {2017})}\BibitemShut {NoStop}%
\bibitem [{\citenamefont {Deng}\ \emph {et~al.}(2015)\citenamefont {Deng},
  \citenamefont {Orgiazzi}, \citenamefont {Shen}, \citenamefont {Ashhab},\ and\
  \citenamefont {Lupascu}}]{Deng_2015_Floquet}%
  \BibitemOpen
  \bibfield  {author} {\bibinfo {author} {\bibfnamefont {C.}~\bibnamefont
  {Deng}}, \bibinfo {author} {\bibfnamefont {J.-L.}\ \bibnamefont {Orgiazzi}},
  \bibinfo {author} {\bibfnamefont {F.}~\bibnamefont {Shen}}, \bibinfo {author}
  {\bibfnamefont {S.}~\bibnamefont {Ashhab}}, \ and\ \bibinfo {author}
  {\bibfnamefont {A.}~\bibnamefont {Lupascu}},\ }\href@noop {} {\bibfield
  {journal} {\bibinfo  {journal} {Phys. Rev. Lett.}\ }\textbf {\bibinfo
  {volume} {115}},\ \bibinfo {pages} {133601} (\bibinfo {year}
  {2015})}\BibitemShut {NoStop}%
\end{thebibliography}%

\end{document}